\documentclass[useAMS,usenatbib]{mn2e}
\usepackage{graphicx}
\usepackage{epstopdf}

\title[LL GRBs and new generation satellites]{Probing the low-luminosity GRB population with new generation satellite detectors}
\author[A. Imerito, D. Coward, R. Burman and D. Blair]{A. Imerito$^{1}$\thanks{E-mail: alan@physics.uwa.edu.au}, D. Coward$^{1}$, R. Burman$^{1}$ and D. Blair$^{1}$\\
$^{1}$School of Physics, University of Western Australia, Crawley, Australia}
\begin{document}

\date{Accepted xxx. Received yyy; in original form zzz;    \textbf{~~~this draft: \today  ~~~Rev S03-01}}

\pagerange{\pageref{firstpage}--\pageref{lastpage}} \pubyear{2008}

\maketitle

\label{firstpage}

\begin{abstract}
We compare the detection rates and redshift distributions of low-luminosity (LL) GRBs localized by {\it Swift} with those expected to be observed by the new generation satellite detectors on {\it GLAST} (now {\it Fermi}) and, in future, {\it EXIST}. Although the {\it GLAST} burst telescope will be less sensitive than {\it Swift}'s in the 15--150 keV band, its large field-of-view implies that it will double {\it Swift}'s detection rate of LL bursts. We show that {\it Swift, GLAST} and  {\it EXIST} should detect about 1, 2 \& 30 LL GRBs,
respectively, over a 5-year operational period. The burst telescope on {\it EXIST} should detect LL GRBs at a rate of more than an order of magnitude greater than that of {\it Swift}'s BAT. We show that the detection horizon for LL GRBs will be extended from $z \simeq 0.4$ for {\it Swift} to $z \simeq 1.1$ in the {\it EXIST} era. Also, the contribution of LL bursts to the observed GRB redshift distribution will contribute to an identifiable feature in the distribution at  $z \simeq 1$.
\end{abstract}


\section{Introduction}
GRBs are the brightest transient astronomical events known. Multi-wavelength observations of GRBs have confirmed that a significant fraction of long GRBs\footnote {Hereafter, `GRB' refers to bursts classified as long.} are associated with the collapse of short-lived massive stars \citep{Hjorth2003, Stanek2003}.  A number of satellites equipped for GRB detection have been placed in low Earth orbit, the most recent of these being the orbiting observatories {\it Swift}$\,$\footnote{http://swift.gsfc.nasa.gov} (launched 2004 November; \citealt{Gehrels2007}) and {\it GLAST}$\,$\footnote{http://glast.gsfc.nasa.gov} (launched 2008 June). Up to the end of 2008 July {\it Swift} has detected 345 bursts\footnote{http://swift.gsfc.nasa.gov/docs/swift/archive/grb\_table.html}  since its launch. In this period 116 GRB redshifts have been measured from this satellite's rapid localization capability and by follow-up spectroscopy of the optical/NIR afterglow and host galaxies by large ground-based telescopes.

Prior to GRB 980425 it was thought that all GRBs were from a limited range of energies \citep{Ramirez-Ruiz2005}. This particular GRB was the first to be associated with a supernova of Type Ib/c and was also the first low-luminosity (i.e., under-luminous) GRB (LL GRB) to be identified \citep{Ramirez-Ruiz2005}. The total (isotropic-equivalent) gamma-ray energy emission was some 5--6 orders of magnitude less than for a typical `classical' GRB (fig. 2 of \citealt{Ghirlanda2006}), demonstrating that GRBs can occur with a wide range of energies \citep{Woosley2004}. Initially it was thought that this GRB was anomalous but more recent detections have challenged that view. GRBs 031203 and 060218 have been identified as possible analogues of GRB 980425 (\citealt{Soderberg2004, Ghisellini2006, Watson2006}) and these three can be considered to form a `typical' set of LL GRBs. For a brief description of these three bursts see \citet{Guetta2007}.


 Because of their sub-energetic nature and the flux limits of recent satellite detectors, the LL GRBs detected thus far are in the local universe ($z  \la  0.1$). Apart from the three typical LL GRBs mentioned above, other subluminous GRBs have been detected --- e.g., GRB 050826 \citep{Mirabal2007} --- some of which have been classified as X-ray flashes --- e.g., XRFs 020903, 030723  and 060218 \citep{Pian2006} --- or X-ray rich --- e.g., GRB 040223 \citep{McGlynn2005}.  X-ray flashes may represent an extension of the GRB population to low energies \citep{Barraud2005}.
 
 Although LL GRB events may be more numerous than `classical' events \citep{Pian2006, Guetta2007, Liang2007} only a few have been observed in the local universe \citep{Soderberg2004, Guetta2007, Daigne2007}. Small-number statistics limit the precision in estimating the local rate density of LL GRBs --- see uncertainties for values of this parameter reported by \citet{Guetta2007} and \citet{Liang2007}.


 We show in this paper that the improved sensitivity and larger field-of-view (FoV) of these new generation satellite detectors will dramatically increase the statistics of these currently rare bursts. We also show how the improved capabilities of these detectors allows detection of LL GRBs at higher $z$.
(For a perspective on studies of  high-$z$ GRBs with new generation instruments see \citet{Salvaterra2008}.)

The aim of this study is to compare the detection capabilities of three gamma-ray detector satellites for observing LL GRBs, focusing on GRB detection rates of the primary burst detectors of {\it Swift}, {\it GLAST} and {\it EXIST}$\,$\footnote{http://exist.gsfc.nasa.gov} (intended to be launched about 2015). We utilize only the primary detector parameters of flux sensitivity, FoV and detector bandwidth for this study. For {\it GLAST} we assume the Gamma-ray Burst Monitor (GBM) is the detector that will trigger on and monitor LL GRBs \citep{Steinle2006}. Similarly, for {\it EXIST} we assume the High Energy Telescope (HET) is the primary LL GRB detector. Thus, our results for {\it GLAST} and {\it EXIST} define a lower bound on the detection capabilities of these new satellite detectors.

\section{Satellite detectors}
\label{sec:SatDetectors}
In the {\it Swift} era the detection rate of LL GRBs is low. To gain an understanding of the progenitors and the mechanisms that cause LL GRBs demands an improved detection rate. The sub-energetic nature of these events requires detectors with increased sensitivity and/or FoV. We show in this paper how the improvement of these two parameters will increase the rate of detection of LL GRBs. We use {\it Swift} as a reference detector against which the newer detectors are compared.

We briefly describe below the detector parameters for the three satellites. In the next section we will compare the predicted observational capabilities of these instruments. Although we restrict the model to the major NASA high-energy observatories we note that other detectors suitable for LL GRB studies are operational or are planned: {\it AGILE}$\,$\footnote{http://agile.asdc.asi.it}(launched 2007 April) is a currently flying detector sensitive in the 30 MeV--50 GeV band; {\it JANUS} is a NASA Small Explorer (SMEX) early universe mission (planned launch 2012) with a 1--20 keV X-ray flash monitor;  {\it MAXI}$\,${\footnote{http://kibo.jaxa.jp/en/experiment/ef/maxi}}(planned launch 2009 May), as an external facility of the International Space Station, will continuously monitor the whole sky for X-rays in the 0.5--30 keV energy band; and {\it SVOM} (planned launch 2012), is a Sino-French GRB satellite detector with a trigger spectral domain of 4--250 keV and a mission design that encompasses rapid ground-based follow-up observations.

On board {\bf{\emph{Swift}}}, bursts are detected by the Burst Alert Telescope, BAT \citep{Barthelmy2005}, a 1.4 sr FoV, 15--150 keV coded mask detector with a 5200 cm$^2$ cadmium-zinc-teluride detector plane. Its peak photon flux sensitivity limit corresponds to $\sim 10^{-8}$ ergs cm$^{-2}$ s$^{-1}$ \citep{Gehrels2004}.

The {\bf Gamma-ray Large Area Space Telescope} ({\it GLAST}$\,$\footnote{Renamed \emph{Fermi} Gamma-ray Space Telescope on 2008-08-26.}), launched in 2008 June, is a current generation satellite for gamma-ray astronomy. The {\it GLAST} GBM consists of 12 NaI detectors for the 8 keV to 1 MeV  range, and two bismuth germanate (BGO) detectors for the 150 keV to 30 MeV range \citep{Carson2007}, with its burst trigger band being 50--300 keV \citep{Band2008, vonKienlin2004}. Its FoV is  9 sr, about 6$\frac{1}{2}$ times larger than {\it Swift}'s.

The {\bf Energetic X-ray Imaging Survey Telescope} ({\it EXIST}) is currently proposed to be a free-flying mission to detect GRBs \citep{Band2007} and conduct a hard X-ray sky survey \citep{Grindlay2003}. {\it EXIST} will have two sets of coded mask telescope systems: a High-Energy Telescope (HET) to cover the 5--600 keV band and a Low-Energy Telescope (LET) for the 3--30 keV band. The (partially-coded) FoV of the HET will be about 2 sr (J. Grindlay, personal communication, 2008-08-06).

To compare detection rates of LL GRBs for these detectors we use their peak flux detection thresholds. Determining the relative detection capabilities of these instruments is not simple for a number of reasons (e.g., see \citealt{Band2002}): the detectors are sensitive across different energy ranges and also vary over the FoV; the triggering algorithms are peculiar to each detector; accumulation times vary and there are synergies and interferences between instruments on the same satellite. As our study is a gross comparison of detector capabilities we will take an uncomplicated approach to quantifying the relative detection capabilities of these three detectors.

In \citet{Band2003}, detection thresholds for the three detectors are normalized to the 1--1000 keV energy band. We approximate the peak flux thresholds of the detectors by using figs. 7,  8 \& 9 of \cite{Band2003}. To approximate the peak flux thresholds we take the midpoint value of the peak flux range as defined by the trigger band of each detector. We obtain (2.0, 3.5, 0.3)   ph~cm$^{-2}$~s$^{-1}$, respectively\footnote {All 3-tuples in this paper denote values for ({BAT, GBM, HET}).}. We express this in terms of relative sensitivities of (1.0, 0.6, 6.7) using  the {\it Swift} BAT as the reference detector. These values are employed in the following section to determine the comparative flux thresholds of these detectors.

\section{GRB DETECTION RATE MODEL}
The aim is to calculate the distribution of detectable GRBs as a function of $z$ using the primary detector parameters mentioned above. The model assumes a flat cosmology with $H_{\mathrm 0}$ = 71 km s$^{-1}$ Mpc$^{-1}$, $\Omega_M$ = 0.7 and $\Omega_\Lambda$ = 0.3  (the concordance cosmology). We employ a GRB detection rate model with two free parameters: the GRB local rate density, $r_0$, and the (two-jet) beaming factor, $f_b^{-1} = (1-\cos\,\theta)^{-1}$, where $\theta$ is the opening half-angle of a jet. We note that both these parameters are uncertain. For simplicity we fix $f_b^{-1}$ and constrain $r_0^{\mathrm {HL}}$ from the model and the observed HL {\it Swift} GRBs.

\subsection{\bf GRB transient event rate model}
To calculate the detection rate of GRBs we start with the following differential rate equation (see \citealt{Coward2007}, \citealt{Coward2001}), which is valid for all independent cosmological transient events:

\begin{equation}
\label{eqn:event_rate}
dR(z)/dz = 4 \pi (c^3 r_0/H^3_0) e(z) F(z, \Omega_{\mathrm M} , \Omega_{\mathrm \Lambda})/(1+z)\;.
\end{equation}
Here $dR/dz$ is the GRB differential event rate in units of $\mathrm s^{-1}$~per unit redshift, $e(z)$ is the dimensionless source rate density evolution function (scaled so that $e(0) = 1$) and $F(z, \Omega_{\mathrm M} , \Omega_{\mathrm \Lambda})$  is a cosmology-dependent function (Eqn. 13.61 of \citealt{Peebles1993}):

\begin{equation}
F(z, \Omega_M , \Omega_\Lambda) = [\;\frac{H_0 D_{\mathrm L}/c}{(1+z)}\;]^2 \;[\;\Omega_{\mathrm M} (1+z)^3 + \Omega_{\mathrm \Lambda}\;]^{-1/2}\;,
\end{equation}
where $D_{\mathrm L}$ is the luminosity distance. For this analysis we take $e(z)$ to follow the SFR model of \citet{Hopkins2006}. Integrating Eqn. \ref{eqn:event_rate} over $z$ gives the cumulative GRB rate.

An `effective' (or potentially observable) differential event rate that takes beaming into account is
 \begin{equation}
 \label{eqn:eff_event_rate}
 dR_{\rm eff}(z)/dz = f_{\mathrm b}~dR(z)/dz\;.
 \end{equation}
It is the GRB rate observed by an infinitely sensitive detector with an unobscured all-sky view.

\begin{table} 
\caption{We employ two parameters in the model: the GRB local rate density, $r_0$, and the beaming factor, $f_b^{-1}$. The {\it Swift} detection rate inferred from the total number of bursts over the mission time constrains $r^{\mathrm {HL}}_0$; $r^{\mathrm {LL}}_0$ is taken from \citet{Guetta2007} and the $f_b^{-1}$ are from \citet{Guetta2005} and \citet{Liang2007}.}
\label{tbl:GRB_params}
\begin{tabular}{ccc}
\hline
 {\bf GRB}       &    {\bf r$_{\mathrm 0}$ }              &   {\bf f$_{\mathrm b}^{-1}$}    \\
 {\bf type}        &    {\bf [Gpc$^{-3}$ yr$^{-1}$]}    &                                                     \\\hline
LL                    &    380                                             &      7                                            \\
HL                   &       15                                             &    75                                            \\
\hline
\end{tabular}
\end{table}


\begin{table} 
\caption{Primary satellite detector parameters that determine the detectability of GRBs. The FoV and trigger band for BAT and GBM are taken from table 1 of \citet{Band2008} and  \citet{Grindlay2008} for {\it EXIST}. The BAT sensitivity is taken from table 2 of \citet{Gehrels2004}. The sensitivities of the other detectors are scaled by this value as noted in Section \ref{sec:SatDetectors}.}
\label{tbl:instr_params}
\begin{tabular}{lccc}
\hline
 {\bf ~Detector}                 &   {\bf Sensitivity (F$_{\rm {lim}}$)}      &    {\bf FoV}     &     {\bf Band \boldmath $[e_1,e_2]$}       \\
 {\bf (Satellite)}                 &   {\bf [ergs s$^{-1}$]}                             &    {\bf [sr]}       &      {\bf [keV] }      \\\hline
 BAT ({\it Swift})                &     $1.0 \times 10^{-8}$                         &    1.4              &      15--150          \\
GBM ({\it GLAST})           &    $1.75 \times 10^{-8}$                        &     9.0              &       50--300        \\
HET ({\it EXIST})              &    $1.5 \times 10^{-9}$                          &     2.0              &      5--600         \\
\hline
\end{tabular}
\end{table}

For the more realistic scenario of a flux-limited detector we employ a single piecewise-smooth GRB LF (luminosity function), $\Phi (L)$, which is comprised of two components, $\Phi_{\rm LL}$ and $\Phi_{\rm HL}$, as a means to account for the LL and HL GRBs. We define LL GRBs as those with burst luminosities $< 10^{49}$ ergs s$^{-1}$. $\Phi_{\rm LL}$, is that part of $\Phi$ defined for LL GRBs  and $\Phi_{\rm HL}$ is for HL GRBs. We adapt the double power-law LF of \cite{Guetta2005} for both, with slopes $-0.1$ and $-2$, respectively. The limits of $\Phi$ were extended to [$5 \times 10^{46}, 3.2 \times 10^{52}$] ergs  s$^{-1}$ in order to encompass the extremely LL GRB 980425 and HL GRBs. The fraction of detectable GRBs, or flux-limited selection function (those observed with peak flux  $> F_{\rm lim}$), for the LF is

\begin{equation}
\label{eqn:flux_select}
\psi_{\rm flux}(z)= \int_{L_{\rm lim}(F_{\rm lim},z)}^{\infty}\phi (L_{\rm iso})~{\rm d}L_{\rm iso}\;,
\end{equation}
where $L_{\rm iso}$ is the isotropic-equivalent luminosity, $F_{\rm lim}$ the lower limit of the detector flux sensitivity and $L_{\rm lim}(F_{\rm lim},z)$ the minimum (isotropic-equivalent) luminosity detectable at a redshift $z$ with detector sensitivity $F_{\rm lim}$. $L_{\rm lim}$ is determined from Eqn.~\ref{eqn:lumin_lim} below. The resulting function, $\psi_{\rm flux}(z)$, represents a dimensionless (detector-dependent) scaling function with range [0,\,1]. The integrand in Eqn.~\ref{eqn:flux_select}, $\phi(L)$, is the normalized $\Phi(L)$:
\begin{equation}
\label{eqn:norm_phi}
\phi(L) \equiv \frac{\Phi (L)}{\int_{0}^{\infty} \Phi (L)\;{\rm d}L}\;.
\end{equation}

\subsection{\bf Instrument model}
We include two correction factors for determining the limiting luminosity for each detector: (i) $B(e_1,e_2)$, the bandpass ratio and (ii)  $k(z)$, the k-correction factor \citep{Hogg2002, Bloom2001}.

(i) $B$ assumes that a detector sees only a window of the GRB gamma-ray spectrum used to define the bolometric isotropic-equivalent luminosity of a $\gamma$-burst:
\begin{equation}
B(e_1,e_2) \equiv  \frac{\int_{e_2}^{e_1} E N(E) {\rm d}E}{\int_{E_2}^{E_1} E N(E) {\rm d}E}\;,
\end{equation}
where the interval $[e_1, e_2]$ is the spectral energy band that the detector is sensitive to and $[E_1, E_2]$ covers the bolometric gamma-ray spectrum that is used for measuring the GRB flux. $N(E)$ is the ensemble photon flux density ($\mathrm {ph\;s^{-1}\;cm^{-2}\;keV^{-1}}$) at the photon energy $E$. The model burst spectrum used here is taken from \cite{Band1993}, with $\alpha = -1$ and $\beta = -2.3$ \citep{Preece2000} for both LL and HL GRBs. We take $E_{\rm peak} = 100$ keV for LL GRBs \citep{Band2002, Kippen2003} and 250 keV for HL GRBs -- fig. 2 of \citet{Amati2006} or fig. 5 of \citet{Preece2000}. For this model we have not adjusted the observed $E_{\rm peak}$ for LL GRBs to enforce conformity with the Amati ($E_{\rm p,i}$--$E_{\rm iso}$) relation \citep{Amati2002}. (See Section \ref{sec:ResultsDiscussion} for further discussion.)

(ii) $k(z)$ accounts for the downshift of $\gamma$-ray energy from the burst to the observer's reference frame and is defined by
\begin{equation}
k(z) \equiv  \frac{\int_{e_2}^{e_1} E N(E) {\rm d}E}{\int_{(1+z)e_2}^{(1+z)e_1} E N(E) {\rm d}E}\;.
\end{equation}
It is obtained by integrating the spectral flux density equation (Eqn. (6) of \citet{Hogg2002} or Eqn. 13.57 of \citet{Peebles1993}) with respect to $\nu_{\rm o}$, noting that $\nu_{\rm e} = (1+z)\nu_{\rm o}$, where the subscripts `o' and `e' identify the observer and emitter reference frames, respectively.

The limiting luminosity as a function of $z$ is determined for each detector from
\begin{equation}
\label{eqn:lumin_lim}
L_{\mathrm {lim}}(z) = 4 \pi D^2_{\mathrm L}(z) F_{\mathrm {lim}}\,B^{-1}(e_1,e_2)\,k(z)\;.
\end{equation}
This is used with $\phi(L)$, via Eqn. \ref{eqn:flux_select}, to derive the selection function, $\psi_{\rm flux }(z)$. We define an instrument effectiveness function by

\begin{equation}
\label{eqn:eff_fn}
\psi_{\rm eff}(z) \equiv \psi_{\rm flux }(z)\, \Omega/4\pi\;,
\end{equation}
where $\Omega$ is the FoV of the detector in steradians; $\psi_{\rm eff}(z)$ is interpreted as the probability of the detector triggering on a single GRB without beaming and at a redshift of $z$.

\subsection{\bf Satellite detection rates}
The model described above can be applied independently to different sub-populations of GRBs. We employ a model for the satellite detection rate that assumes two different GRB populations with normalized LFs, $\phi_{\rm LL}$ and $\phi_{\rm HL}$ for the LL and HL bursts respectively. The detection rate of a GRB population i for a detector is given by
\begin{equation}
\label{eqn:popn_obs_rate}
dR_{\rm obs}^{\rm (i)}(z)/dz = \psi_{\rm eff}^{\rm (i)}(z)\,.\,dR_{\rm eff}^{\rm (i)}(z)/dz\;,
 \end{equation}
with the total GRB differential event rate the sum over GRB populations (in this study, i =  LL, HL):
\begin{equation}
\label{eqn:tot_obs_rate}
d{\mathcal R}_{\rm obs}(z)/dz = \sum_{\rm i}dR_{\rm obs}^{\rm (i)}(z)/dz\;.
 \end{equation}
Table~\ref{tbl:GRB_params} lists the parameters used to calculate the effective LL and HL GRB event rates (via Eqn.~\ref{eqn:eff_event_rate}). Table~\ref{tbl:instr_params} lists the detector parameter values used to determine the detection rates of GRBs (via Eqn.~\ref{eqn:popn_obs_rate}).
 
\section{Results \& Discussion}
\label{sec:ResultsDiscussion}
\begin{figure}
\includegraphics[scale=0.40]{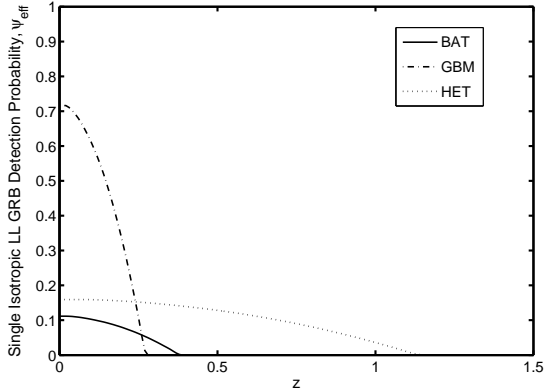}
\caption{The LL GRB detection probability, $\psi_{\rm eff}$, for the 3 satellite detectors. The curves show the relative detection efficiencies for detecting LL GRBs. The functions are determined by $F_{\rm lim}$, the k-correction, bandwidth scaling, FoV and LF. As the LF is identical for all detectors, differences between the curves result from instrument parameters. At $z = 0$, the probability of detecting a single isotropic LL GRB is just the detector's FoV (as a sky-fraction). The detection horizon distance for the detector, $z_{\rm h}$, is defined by $\psi_{\rm eff}(z_{\rm h}) = 0$. The detection horizon is $z \le 1.1$ for all three detectors. The high sensitivity of HET results in a detection horizon which is about 3--4 times that of the other two detectors.}
\label{fig:DetectorScFns}
\end{figure}

We have not assumed that the $E_{\rm p,i}$--$E_{\rm iso}$ correlation is valid for LL GRBs. While GRB 060218 obeys this relation the other LL GRBs (980425 and 031203) do not (fig. 3 of \citet{Amati2006}). Although \citet{Ghisellini2006} have indicated that the two outliers may indeed obey the Amati relation if spectral evolution is taken into account there appears to be no compelling physical evidence to support this at present. Given the above, we take a conservative approach, retaining the Band spectrum and setting $E_{\rm peak} = 100$ keV for LL GRBs (which is between the observed values for GRB 980425 and 031203 -- see fig. 3 or tables 1 \& 2 of \citet{Amati2006}). Calculations made with $E_{\rm peak} = 10$ keV reduce the horizon limits (see next paragraph) to (85, 70, 90)\% those for $E_{\rm peak} = 100$ keV. Similarly, detection rates are (60, 35, 75)\% those of the higher $E_{\rm peak}$. However, these do not change the qualitative results we report here. With greater numbers of LL GRBs detected in future the observed $E_{\rm peak}$ may be used to confirm whether LL GRBs are consistent with the Amati relation or not.

The calculated detection probabilities for the three satellites for LL GRBs are shown in Fig. \ref{fig:DetectorScFns}. We define the `horizon distance', $z_{\mathrm h}$, of a detector to be the greatest distance at which a LL GRB can trigger the detector ($z_{\mathrm h} \propto \sqrt{F_{\rm lim}}$).

\begin{figure}
\includegraphics[scale=0.40]{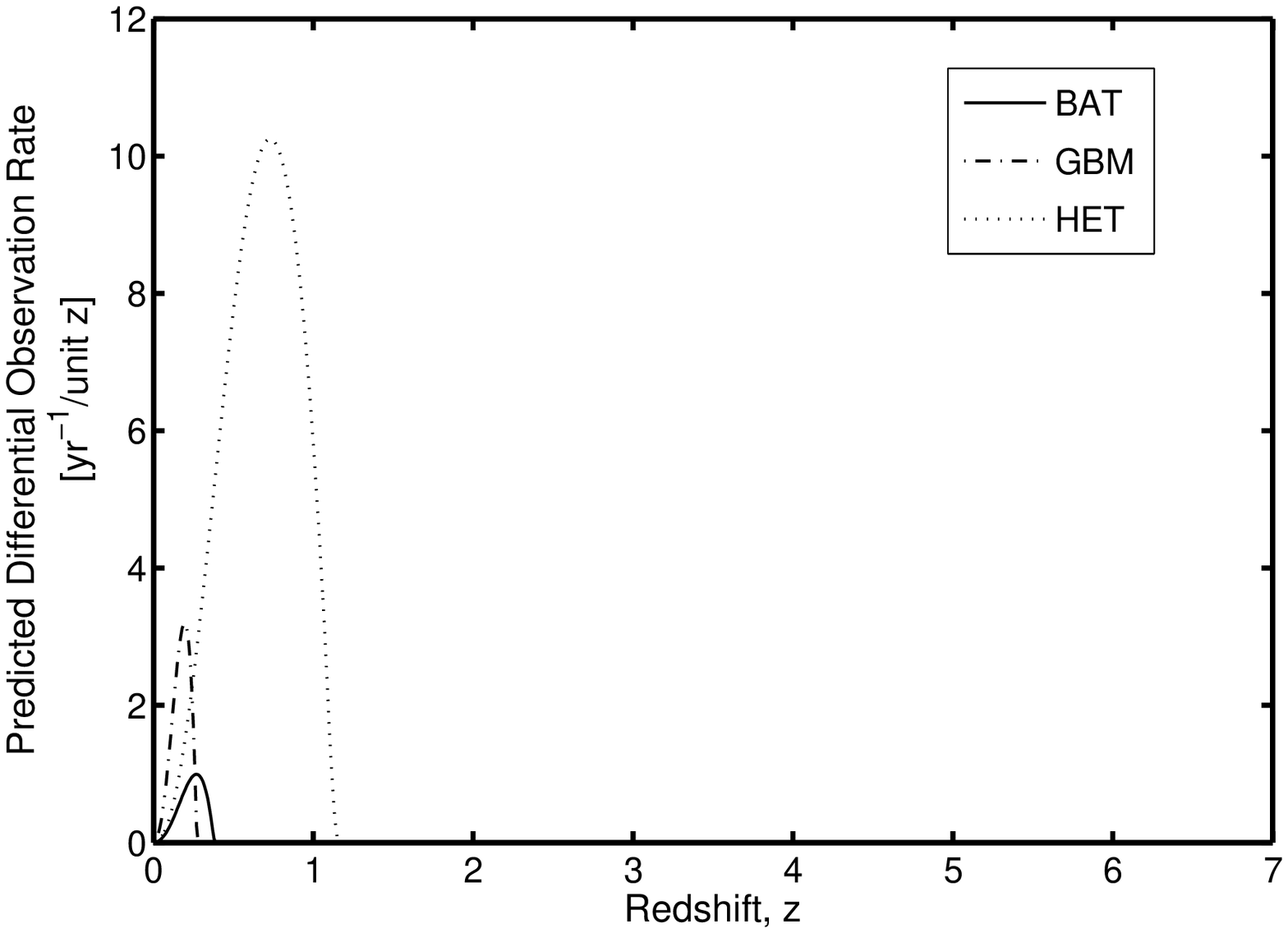}
\includegraphics[scale=0.40]{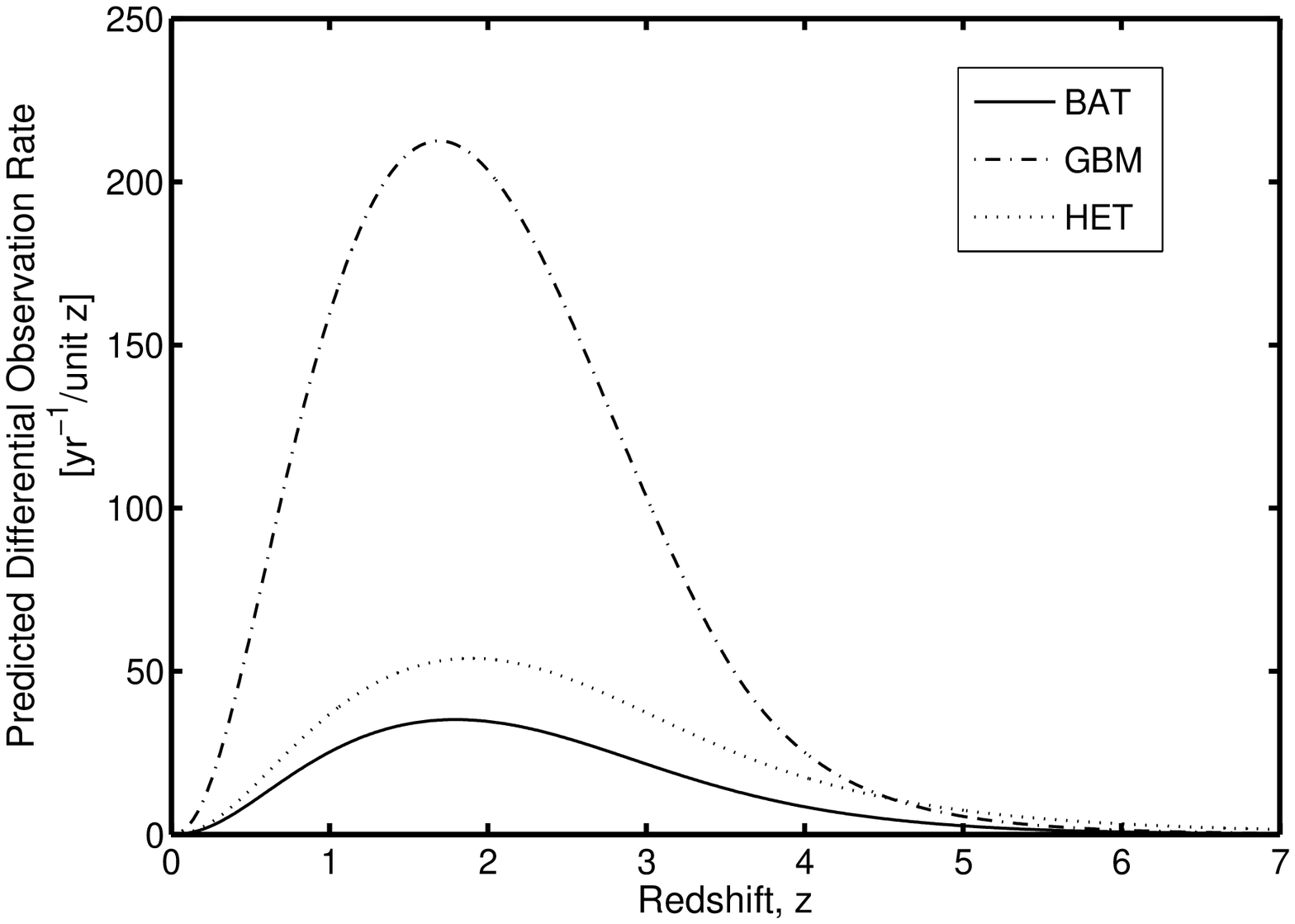}
\caption{Predicted differential observation rates for BAT, GBM and HET for LL (upper panel) and HL GRBs (lower panel). The GRB rate density peak values for GRBs  are determined by the sensitivity and FoV of the detectors while their $z$-locations are determined by sensitivity (and, to a lesser extent, redshift).}
\label{fig:PredObsRates_LLandHL}
\end{figure}

\begin{figure}
\includegraphics[scale=0.40]{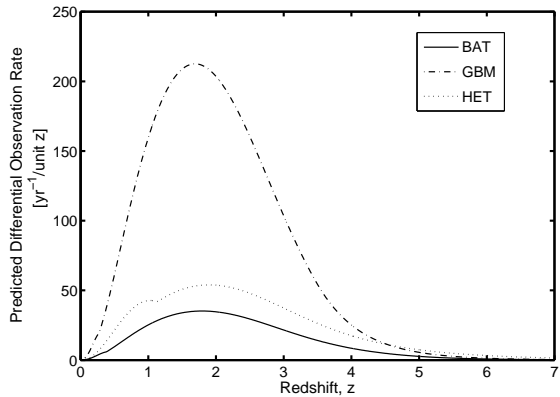}
\caption{Predicted differential detection rates for the GRB population comprised of LL \& HL bursts. These curves are the sum of the LL \& HL detection rate models (Fig. \ref{fig:PredObsRates_LLandHL}). The effect of two different beaming factors results in a LL GRB `notch' at ${z \approx 1}$ for HET. Because of the lower sensitivities of the GBM \& BAT this feature is not visible to those detectors.}
\label{fig:PredObsRates_Combined}
\end{figure}

We convolve the instrument effectiveness above with the GRB beaming factor and GRB rate evolution, $e(z)$, to produce detection rate curves (i.e., $dR_{\rm obs}(z)/dz$) for each detector and GRB population. The predicted distributions of LL and HL GRB detections are plotted in Fig. \ref{fig:PredObsRates_LLandHL}. The combined GRB detection  rate curves are shown in Fig. \ref{fig:PredObsRates_Combined} for each detector. We note that the LL GRB distribution produces a noticeable `notch' at $z \approx 1$ for HET, a result of its higher sensitivity.

\begin{table} 
\label{tbl:results}
\caption{Predicted LL GRB detection rates and horizons.}
\begin{tabular}{lcc}
\hline
{\bf ~Detector}               &   {\bf LL GRB}          &    {\bf LL Horizon}                    \\
{\bf (Satellite)}                &   {\bf [yr$^{-1}$]}     &    {\bf [z$_{\mathrm h}$] }        \\\hline
BAT ({\it Swift})              &     0.2                         &      0.4                                         \\
GBM (${\it GLAST}$)    &     0.4                        &       0.3                                         \\
HET (${\it EXIST}$)       &     6.5                        &      1.1                                          \\
\hline
\end{tabular}
\end{table}

The locations of the LL GRB peaks in redshift (upper panel of Fig.~\ref{fig:PredObsRates_LLandHL}) is determined by the relative sensitivities (and, to a lesser extent, trigger bandwidths) of the detectors. A more sensitive detector will be triggered by a greater number of distant GRBs. The peak LL detection rates of BAT and GBM occur in the redshift regime of $[0.15, 0.3]$ while for HET they occur at the significantly higher redshift of just over 0.7. The LL GRB detection horizon distances of the above detectors are $z = (0.4, 0.3, 1.1)$. The sensitivity of HET will also enable it to detect HL GRBs at higher $z$, thus probing a larger volume of space, as shown by the lower panel of Fig.~\ref{fig:PredObsRates_LLandHL}. 

The height of each curve in Fig. \ref{fig:PredObsRates_LLandHL} is proportional to the FoV of the detector. For a bright GRB population the FoV is the dominant parameter, but sensitivity becomes important for detecting fainter bursts, as illustrated by comparing the GBM and HET curves for both populations in Fig.~\ref{fig:PredObsRates_LLandHL} (upper and lower panels). The moderately large FoV of HET coupled with its relatively high sensitivity implies it will detect an order of magnitude more LL GRBs than the other two satellites combined (Table~3). The large FoV of GBM enables it to capture the most bursts for $z \la 0.3$ even though it is the least sensitive detector in this comparison study. Its large FoV also makes it the optimal detector out of the three for HL GRBs out to redshift about 4.5. Fig. \ref{fig:PredObsRates_Combined} indicates that HET will also be able to detect GRBs in the early universe, thus potentially probing the evolutionary history of GRBs.

From the upper panel of Fig.~\ref{fig:PredObsRates_LLandHL} we note that HET will detect large numbers of nearby LL GRBs and, thus, significantly add to the statistics of this under-luminous population. In addition, Fig.~\ref{fig:DetectorScFns} shows that HET takes over from GBM as the most efficient LL GRB detector for $z \ga 0.3$ in spite of it having a quarter of the GBM's FoV. This is due to HET's higher sensitivity, which is also reflected in HET having a much larger LL horizon distance than the other two detectors.

{\it GLAST}$\,$'s GBM, having the largest FoV, produces thebroad and high differential detection rate prominent in the lower panel of Fig.~\ref{fig:PredObsRates_LLandHL}. The predicted $z$-integrated detection rates of LL GRBs are (0.2, 0.4, 6.5) yr$^{-1}$ for BAT ({\it Swift}), GBM ({\it GLAST}) and HET ({\it EXIST}), respectively.

The notch in Fig.~\ref{fig:PredObsRates_Combined} for the HET curve is a result of two factors: (1)  HET's high sensitivity will detect a higher proportion of LL GRBs than will the other two satellites and (2) the low LL GRB beaming factor and high local rate density (compared to HL GRBs) results in an effective differential LL rate about 250 times that of HL GRBs. However, it may take two to three decades for this feature to become apparent. Hence, it is unlikely that any single satellite observatory will detect this `hidden' LL peak during an operational career. Nonetheless, even the 3 or so LL GRBs detectable by {\it Swift} and {\it GLAST} in the next 5 years will be a significant addition and will provide constraints on this population.

We note two selection effects that may be significant but are not included in this study. Firstly, as \citet{Band2006} showed, the BAT is more sensitive to high-redshift bursts because of their longer duration. Secondly, the rate density of GRBs may be evolving at high redshift, as suggested by \citet{Daigne2006} and \citet{Le2007}. \citet{Firmani2004} reach a similar conclusion, but explain a higher than expected rate as evidence for an evolving GRB luminosity function. These issues remain uncertain. However, their effect does not influence the main results from this study since our focus is on the detectability of small-redshift LL bursts.

We also note that selection effects that plague the present GRB redshift distribution need to be understood to exploit fully data for studying GRB rate evolution. \citet{Coward2008} show that selection effects are especially dominant in the optical afterglow and spectroscopic observations of {\it Swift}-triggered bursts.

In summary, we show that HET will probe both LL GRBs and early-universe GRBs while GBM will be more sensitive to the HL population out to $z \approx 4.5$. Surprisingly, we find that despite the lower sensitivity of GBM, its large FoV implies a LL GRB detection rate greater than that of BAT and HET out to the GBM's sensitivity horizon of $z = 0.3\;.$ Importantly, HET will increase the LL GRB detection rate by a factor of about 30 compared to {\it Swift}'s BAT. It will open up a new detection regime by probing the LL GRB population in a significantly larger volume out to $z \sim 1$.

\section{Acknowledgements}
We thank the referee for providing several suggestions that have helped clarify the results.\\

A. Imerito is supported by the Australian Research Council (ARC) grant LP0667494. D. M Coward is supported by ARC grants DP0877550, LP0667494 and the University of Western Australia.


\label{lastpage}

\newpage


\begin{thebibliography}{99}

\bibitem[Amati et al. (2002)]{Amati2002} Amati L. et al., 2002, A\&A, 390, 81

\bibitem[Amati (2006)]{Amati2006} Amati L., 2006, MNRAS, 372, 233

\bibitem[Band (2002)]{Band2002} Band D.L., 2002, ApJ, 578, 806

\bibitem[Band (2003)]{Band2003} Band D.L., 2003, ApJ, 588, 945

\bibitem[Band (2006)]{Band2006} Band D.L., 2006, ApJ, 644, 378

\bibitem[Band (2007)]{Band2007} Band D.L., 2007, arXiv:0710.4602v1 [astro-ph]

\bibitem[Band (2008)]{Band2008} Band D.L., 2008, arXiv:0801.4961v1  [astro-ph]

\bibitem[Band et al. (1993)]{Band1993} Band D. et al., 1993, ApJ, 413, 281

\bibitem[Barraud et al.  (2005)]{Barraud2005} Barraud C., Daigne F., Mochkovitch R., Atteia J.L., 2005, A\&A, 440, 809

\bibitem[Barthelmy (2005)]{Barthelmy2005} Barthelmy S. et al., 2005, Space. Sci. Rev., 120, 143

\bibitem[Bloom, Frail \& Sari (2001)]{Bloom2001} Bloom J.S., Frail D.A. and Sari R., 2001, ApJ, 121, 2879

\bibitem[Carson (2007)]{Carson2007} Carson J., 2007, J. of Physics: Conf. Series 60, 115



\bibitem[Coward (2007)]{Coward2007} Coward D.M.,  2007, New Astron. Rev., 51, 539
  
\bibitem[Coward, Burman \& Blair (2001)]{Coward2001} Coward D.M.,  Burman R.R. and Blair D.G., 2001, MNRAS, 324, 1015

\bibitem[Coward et al. (2008)]{Coward2008} Coward D.M., Guetta D., Burman R.R., Imerito A., 2008, MNRAS, 386, 111

\bibitem[Daigne, Rossi \& Mochkovitch (2006)]{Daigne2006} Daigne F., Rossi E.M. and Mochkovitch R., 2006, MNRAS, 372, 1034

\bibitem[Daigne \& Mochkovitch (2007)]{Daigne2007} Daigne F. and Mochkovitch R.,  2007, A\&A, 465, 1

\bibitem[Firmani et al. (2004)]{Firmani2004} Firmani C., Avila-Reese V., Ghisellini G., Tutukov A.V., 2004, ApJ, 611, 1033

\bibitem[Gehrels et al. (2004)]{Gehrels2004} Gehrels N. et al., 2004, ApJ, 611, 1005

\bibitem[Gehrels, Cannizo \& Norris (2007)]{Gehrels2007} Gehrels N., Cannizzo J.K. and Norris J.P., 2007, New J. Phys., 9, 37

\bibitem[Ghirlanda, Ghisellini \& Firmani (2006)]{Ghirlanda2006} Ghirlanda G., Ghisellini G. and Firmani C., 2006, New J. Phys., 123, 1

\bibitem[Ghisellini et al. (2006)]{Ghisellini2006} Ghisellini G., Ghirlanda G., Mereghetti S., Bosnjak Z., Tavecchio F., 2006, MNRAS, 372, 1699


\bibitem[Grindlay et al. (2003)]{Grindlay2003} Grindlay J.E., Craig W.W., Gehrels N., Harrison F.A., Hong J., 2003, Proc. SPIE, 4851, 331

\bibitem[Grindlay (2008)]{Grindlay2008} Grindlay J. (2008), Hard X-Ray/IR Spectral-Imaging of GRBs From the High-z Universe to {\it EXIST}, [Online], Available from: $<$http://www.cospar-assembly.org/abstractcd/COSPAR-08/
 abstracts/data/pdf/abstracts/E18-0005-08.pdf$>$ [2008-08-14]

\bibitem[Guetta, Piran and Waxman (2005)]{Guetta2005} Guetta D., Piran T. and Waxman E., 2005, ApJ, 619, 412

\bibitem[Guetta \& Della Valle (2007)]{Guetta2007} Guetta D. and Della Valle M.,  2007, ApJ, 657, L73

\bibitem[Hjorth et al. (2003)]{Hjorth2003} Hjorth J. et al., 2003, Nat, 423, 847

\bibitem[Hogg et al. (2002)]{Hogg2002} Hogg D.W., Baldry I.K., Blanton M.R. Eisenstein D.J., 2002, astro-ph/0210394v1

\bibitem[Hopkins \& Beacom (2006)]{Hopkins2006} Hopkins A.M. and Beacom J.F., 2006, ApJ, 651, 142

\bibitem[Le \& Dermer (2007)]{Le2007} Le T. and Dermer C.D., 2007, ApJ, 661, 394

\bibitem[Liang et al. (2007)]{Liang2007} Liang E., Zhang B., Virgili F., Dai Z.G., 2007, ApJ, 662, 1111

\bibitem[Kippen et al. (2003)]{Kippen2003} Kippen R.M., Woods P.M., Heise J., in't Zand J.J.M., Briggs M.S., Preece R.D., 2003, AIP Conf. Proc., 662, 244

\bibitem[McGlynn et al. (2005)]{McGlynn2005} McGlynn S. et al., 2005, Il Nuovo Cimento C, 28, 4, 481

\bibitem[Mirabal, Halpern \& O'Brien (2007)]{Mirabal2007} Mirabal N.,  Halpern J.P. and O'Brien P.T., 2007, ApJ, 661, L127


\bibitem[Peebles (1993)]{Peebles1993} Peebles P.J.E., 1993. Principles of Physical Cosmology. Princeton University Press, Princeton, NJ.

\bibitem[Pian et al. (2006)]{Pian2006} Pian E. et al., 2006, Nat, 442, 1011

\bibitem[Preece et al. (2000)]{Preece2000} Preece R.D., Briggs M.S., Mallozzi R.S., Pendleton G.N., Paciesas W.S., Band D.L., 2000, ApJS, 126, 19

\bibitem[Ramirez-Ruiz et al. (2005)]{Ramirez-Ruiz2005} Ramirez-Ruiz E., Granot J., Kouveliotou C., Woosley S.E., Patel S.K., Mazzali P.A., 2005, ApJ, 625, L91

\bibitem[Salvaterra et al. (2008)]{Salvaterra2008}  Salvaterra R., Campana S., Chincarini G., Covino S. and
 Tagliaferri G., 2008, MNRAS, 385, 189


\bibitem[Soderberg et al. (2004)]{Soderberg2004} Soderberg A.M. et al., 2004, Nat, 430, 648

\bibitem[Stanek et al. (2003)]{Stanek2003} Stanek K.Z. et al., 2003, ApJ, 591, L17

\bibitem[Steinle et al. (2006)]{Steinle2006} Steinle H. et al., 2006, Chin. J. Astron. Astrophys., 6, Suppl. 1, 365


\bibitem[von Kienlin et al. (2004)]{vonKienlin2004} von Kienlin A. et al., 2004, Proc. SPIE, 5488, 763  

\bibitem[Watson et al. (2006)]{Watson2006} Watson S.A. et al., 2006, ApJ, 636, 967

\bibitem[Woosley (2004)]{Woosley2004}  Woosley S., 2004, Nat, 430, 623

\bsp

\end{thebibliography}
\end{document}